# INVESTIGATION OF LIGHT NUCLEUS CLUSTERING IN RELATIVISTIC MULTIFRAGMENTATION PROCESSES

M. I. Adamovich[1], V. Bradnova[2], M. M. Chernyavsky[1], V. A. Dronov[1], S. G. Gerasimov[1], L. Just[3], M. Haiduc[4], S. P. Kharlamov[1], K. A. Kotel'nikov[1], A. D. Kovalenko[2], V. A. Krasnov[2], V. G. Larionova[1], F. G. Lepekhin[5], A. I. Malakhov[2], G. I. Orlova[1], N. G. Peresadko[1], N. G. Polukhina[1], P. A. Rukoyatkin[2], V. V. Rusakova[2], N. A. Salmanova[1], B. B. Simonov[5], S. Vokál[2,6], P. I. Zarubin[2]

The BECQUEREL Collaboration

[1] P. N. Lebedev Physical Institute RAS, Moscow, Russia (FIAN)
[2] Joint Institute for Nuclear Research, Dubna, Russia (JINR)
[3] Institute of Experimental Physics SAS, Košice, Slovakia
[4] Institute of Space Sciences, Bucharest-Magurele, Romania
[5] Petersburg Institute of Nuclear Physics, Gatchina, Russia
[6] P. J. Šafárik University, Košice, Slovakia

Abstract

The use of emulsions for studying nuclear clustering in light nucleus fragmentation processes at energies higher than 1A GeV is discussed. New results on the topologies of relativistic $^7$Li and $^{10}$B nucleus fragmentation in peripheral interactions are given. A program of research of the cluster structure in stable and radioactive nuclei is suggested.

Progress achieved in the study with relativistic nucleus beams gives rise to new approaches in solving some topical problems of the nuclear structure. Among them is a search for collective degrees of freedom in which separate groups of nucleons behave like composing clusters. Such a peculiar feature, clustering in excited nuclei, is revealed especially clearly in light nuclei, in which the possible number of cluster configurations is rather small. The natural components of such a picture are few-nucleon systems having no proper nuclear excitations. First of all these are α-particles, as well as pairing proton and neutron states, deuterons, tritons, and $^3$He nuclei. Possibly, the study of the decays of stable and radioactive nuclei to cluster fragments might reveal some new particularities of their origin and their role in cosmic-ray nucleosynthesis.

In our case the use of nuclear beams of energy above 1A GeV is based on the established phenomenon of limiting fragmentation of nuclei. This implies that the isotopic compound of the fragments of a projectile in a narrow forward angular cone is independent of the type of a target-nucleus and the nature of a reaction. The reaction takes up the shortest time. One of the technical merits is the absence of the energy threshold for detecting a fragmentation process.

The most advantageous way for studying clustering is the use of peripheral interactions of relativistic nuclei which occur at a minimal mutual excitations of colliding nuclei caused by electromagnetic and diffractive interactions and the absence of charged meson production. In this case, a clear separation of nuclear fragmentation products according to rapidity is achieved. The requirements of conservation of the electric charge and mass number of a projectile, and narrow angular correlations of relativistic fragments might be employed in the analysis.

The reliable and complete observation of the multiparticle relativistic fragmentation processes is a motivation for which we have used nuclear emulsion technique. Emulsions enable us to establish the most feasible charge channels of such processes. Measurements of multiple scattering angles make it possible to determine the total momentum of hydrogen and helium relativistic fragments and thereby to determine their mass. The record angular resolution of emulsions allows one to reconstruct the invariant mass (excitations) of a fragmenting system.

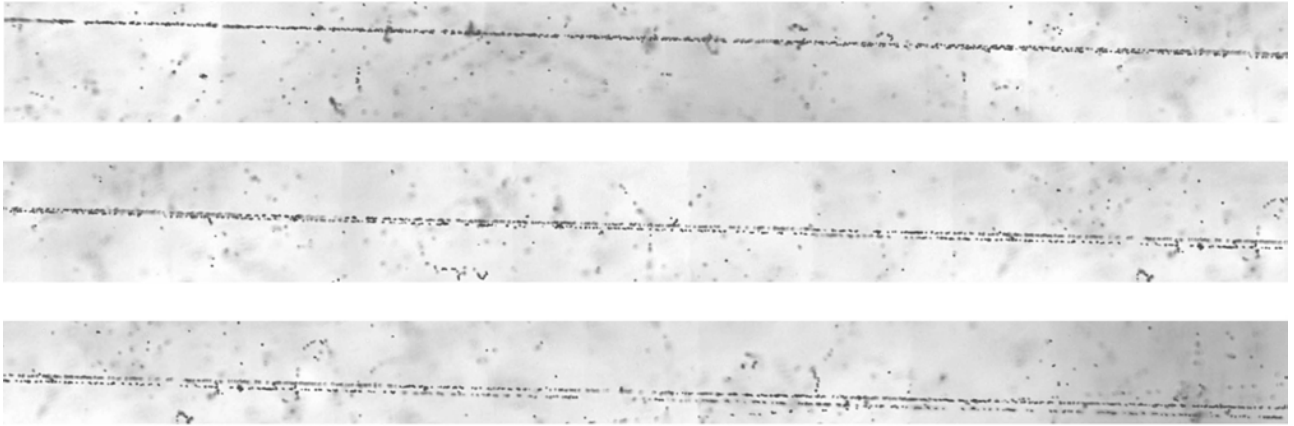

Photograph. Event of coherent $^{10}$B nucleus dissociation to a triple-charged (top) and double charged fragments (bottom) on three consecutive sections of the tracks. A three-dimensional image of the event is reconstructed as a plane projection by means of an automatic microscope of FIAN and the PAVICOM complex.

These considerations have underlain our experimental program, named the BECQUEREL project, aimed at a systematic study of the fragmentation channels of stable and radioactive nuclei at the JINR Nuclotron beams using the emulsion technique. The study of cluster fragmentation of the nuclei $^6$Li ($\alpha$-d) [2-5], $^{12}$C ($3\alpha$) [6-9] and $^{16}$O ($4\alpha$) [10] performed by the emulsion technique serves as a guide for our project. In what follows we discuss the results of research of the $^{10}$B and $^7$Li cluster structure. These results are the basis for further investigations of clustering in light neutron-deficient and heavier stable nuclei.

**$^{10}$B clustering.** $^{10}$B nuclei were accelerated at the JINR Nuclotron, and a $^{10}$B nucleus beam of energy of 1A GeV was formed. The beam was used to irradiate stacks composed of BR-2 type emulsion layers 550 micron thick and measuring $10 \times 20$ cm$^2$ which were sensitive to minimum ionization of single-charged particles. During irradiation, the emulsion layers were located in parallel to the beam direction so that the beam particles could enter the butt-end of the emulsion layers. Search for nucleus-nucleus interactions was performed by visual scanning of particle tracks by means of microscopes with magnification ×900. At a scanned track length of 138.1 m, it was found 960 inelastic interactions of $^{10}$B nuclei. The mean free path of $^{10}$B nuclei to an inelastic interaction in emulsion was found to be 14.4±0.5 cm. This value meets well the dependence of the mean free path upon the atomic number of a bombarding nucleus for light nuclei having homogeneous nucleonic density.

Information about the charge composition of charged fragments and about the channels of $^{10}$B nucleus fragmentation in peripheral collisions has been obtained. We attribute to the peripheral interactions events in which the total charge of relativistic fragments is equal to the charge of the primary $^{10}$B nucleus, the production of charged mesons is not observed, but the production of slow nuclear fragments can occur. In order to single out such events by visual observation we estimated the charges of relativistic particles and the total charge of relativistic particles with emission angles less than 15º with respect to the $^{10}$B direction. For the primary beam energy of 1A GeV, this value corresponds to the proton transverse momentum of 0.44 GeV/c. Then using measuring microscopes we evaluated the emission angles of all particles in the selected events. The particle charges were determined by the length of spacings in their tracks. The number of the found events in which the total charge of fragments is equal to five and in which charged mesons are not observed is equal to 93 (10% of all the events). We notice that the selection of events in which the production of even nuclear emulsion fragments is forbidden decreases statistics down to 41 events. In this case, the distribution of statistics by channels remains practically unchanged.

In 65% of such peripheral interactions the $^{10}$B nucleus is disintegrated to two double charged and a one single-charged particles. A single-charged particle is the deuteron in 40% of these events. 10% of events contain fragments with a charge equal to 3 and 2 (Li and He isotopes), and 2% of events contain a fragment with charges equal to 4 and 1 (Be nucleus and the proton). The $^6$Li production accompanied by an alpha particle may be considered as a correlation of α-particle and deuteron clusters. The photography shows an example of a two-particle decay to Li and He fragments. The fragmentation channel containing α-particle and three single-charged fragments (disintegration of one of the α-clusters) makes up 15%.

An equal correlation of the channels (2He+d)/(2He+p)≈1 is analogous to the $^6$Li fragmentation where (He+d)/(He+p)≈1. These ratios point to an abundant yield of deuterons in the $^{10}$B case too [2,3]. Thus, the deuteron cluster manifests itself directly in the three-particle decays of $^{10}$B nuclei accompanied by two two-charged particles. Another indication to deuteron clustering is a small mean transverse momentum of deuterons $P_t$=0.14±0.01 GeV/c in these events, in just the same way as in the case of $^6$Li, where $P_t$= 0.13 ±0.02 GeV c.

We note that $^{10}$B nucleus, like the deuteron, and $^6$Li and $^{14}$N belong to a rare class of odd-odd stable nuclei. Therefore, it is interesting to establish the presence of deuteron clustering in relativistic $^{14}$N fragmentation.

**$^7$Li clustering.** A total of 1274 inelastic interactions were found to be occurred in a nuclear emulsion irradiated by a $^7$Li beam with a momentum 3A GeV/c at the JINR Synchrophasotron , at a length of 185 m of scanned tracks. The mean free path of $^7$Li nuclei up to an inelastic interaction in emulsion was found to be λ= 14.5±0.4 cm which coincides, within errors, with the $^6$Li mean free path [2,3]. The close values of the mean free paths and the total transverse cross sections of inelastic interactions of $^6$Li and $^7$Li point out that their effective interaction radii are also close in magnitude to each other.

About 7% of all inelastic interactions of $^7$Li nuclei are peripheral interactions (92 events), which contain only the charged fragments of a relativistic nucleus, they do not contain any other secondary charged particles, and the total charge of the fragments is equal to the charge of a fragmenting nucleus. The 80 events are actually two-particle $^7$Li decays to one double and one single-charged fragments. Half of these events are attributed to a decay of $^7$Li nucleus to α-particle and a triton (40 events). The number of decays accompanied by deuterons makes up 30%, and by protons – 20%. The isotopic composition of decayed particles points to the fact that these events are related to the structure as α-particle and the triton clusters. The predominance of tritons in the isotopic compound of single-charged fragments well shows the dominating role of the triton cluster in the $^7$Li fragmentation in very peripheral interactions with emulsion nuclei.

Earlier, similar two-particle $^6$Li decays to α-particle and a deuteron which reflected the weakly bound two-cluster nuclear structure were registered in inelastic peripheral interactions of $^6$Li nuclei with a momentum 4.5A GeV/c in emulsion. Thus, the structure in the form of α particle core and external nucleons bound into a cluster is typical not only of $^6$Li nucleus, but also of $^7$Li one. The obtained value of the cross section for coherent decay to α particle and a triton (27±4 mb) was found to be about the same as the cross section of paper [2] for $^6$Li decay to α particle and a deuteron (22±4 mb). This may be viewed of as an indication to the fact that the mechanisms of the decays in question are of the same nature. It is interesting to continue to clear up a possible role played by the tritons as cluster elements in $^{11}$B and then in $^{15}$N nuclei.

**Clustering with $^3$He participation.** The nucleus $^3$He is a natural element of the cluster picture of excitations of light neutron-deficient nuclei such as $^6$Be, $^7$Be, $^8$B, $^9$C, $^{10}$C, $^{11}$C, $^{12}$N, and heavier ones. If we replace α particle clusters in nuclei $^8$Be, $^9$Be, $^{10}$B, $^{12}$C, and $^{14}$N by the $^3$He nuclei we can obtain similar cluster states. In this approach $^6$Be nucleus is a weakly coupled $^3$He-$^3$He resonance close in its properties to the α-α system in $^8$Be.

By analogy with $^9$Be nucleus, in $^7$Be there are possible n-$^6$Be and $^3$He-n-$^3$He excitations in addition to the α-$^3$He state. It is interesting to separate a $^3$He-$^3$He-$^3$He state in $^9$C as an analogy of α-α-α clustering in $^{12}$C and compare the intensity of its excitation with p-$^8$Be and p-α-α options.

We mention other interesting states like p-p-α-α and α-$^3$He-$^3$He in $^{10}$C, α-α-3He in $^{11}$C, and α-$^8$B in $^{12}$N. The existence of such nuclear-molecular quantum states may points to some alternative scenarios of nucleosynthesis of light nuclei .It might occur via intermediate radioactive nuclei in burning processes of compound isotopic mixtures of hydrogen and helium on the basis of fusion reactions, including simultaneous fusion of a few particles from an intermediate bound state.

It seems to us that the use of emulsions at relativistic radioactive nucleus beams in the domain of light neutron-deficient isotopes is mostly justified. Thanks to the most complete observation, the most significant decay channels for excited nuclei can be established by the charge of the final states. For these channels, it is possible to analyze the mass and angular spectra, to look into the correlations and to estimate specific excitation energies.

As the first step along these lines, we have performed irradiation of emulsions at the JINR Nuclotron. A secondary beam containing a large fraction of $^7$Be nuclei was used. The beam was formed by tuning a magnet-optical channel for optimum choice of the products of accelerated $^7$Li and $^7$Be charge exchange reaction. The cross section of this reaction is of the order of $10^{-4}$ of the inelastic cross section. At present, the results obtained are being analyzed.

We expect that the charge exchange reactions would enable us to form secondary beams by the following charge exchange reactions $^{10}$B→$^{10}$C, $^{11}$B→$^{11}$C, and $^{12}$C→$^{12}$N. $^8$B beam is supposed to be formed via the fragmentation reaction $^{10}$B→$^8$B. We estimate its probability at the available Nuclotron energy at a level $10^{-3}$ of the inelastic cross section. The estimate is grounded on observation of two events of boron nucleus scattering accompanied by high momentum recoils of the target fragments. This possibility deserves careful verification by means of spectrometric measurements. $^9$C nucleus irradiation is the most problematic one because the accompanying $^3$He nuclei possessing the same magnetic rigidity are an unavoidable background.

**Clustering with $^4$He participation.** In this framework, we renewed the analysis of emulsion irradiations by $^{22}$Ne, $^{24}$Mg, $^{28}$Si, and $^{32}$S nuclei at a 4.5A GeV/c momentum. We are planning to search for and investigate fragmentation of these nuclei by observing the final states containing a few α-particles. Search for states considered as nuclear molecules is of special interest. Our approach will enable us to make the choice between the definition of this resonance as a configuration of a few bound α-particles or a resonance arising only in a nuclear scattering. We may hope that the application of the picture of α-particle clustering in nuclei together with peripheral clustering of deuterons, tritons, $^3$He and nucleon pairs will become still wider. The observation of the processes of relativistic multiparticle fragmentation provides the experimental basis for the development of the cluster models of the nucleus.

This work has been supported by the grants of the Russian foundation of fundamental researches N96-15-96423 (Academician A.M.Baldin's scientific school) and N02-02-164 12a, VEGA N1/9036/22 of the Agency of Science of the Ministry of Education of the Slovak Republic and the Slovak Academy of Sciences, and the grants of the JINR Plenipotentiaries of Slovakia and Romania in the year 2002.